\pdfoutput=1
\documentclass[letterpaper]{article} % DO NOT CHANGE THIS
\usepackage[submission]{aaai24}  % DO NOT CHANGE THIS
\usepackage{times}  % DO NOT CHANGE THIS
\usepackage{helvet}  % DO NOT CHANGE THIS
\usepackage{courier}  % DO NOT CHANGE THIS
\usepackage[hyphens]{url}  % DO NOT CHANGE THIS
\usepackage{graphicx} % DO NOT CHANGE THIS
\usepackage{natbib}  % DO NOT CHANGE THIS AND DO NOT ADD ANY OPTIONS TO IT
\usepackage{caption} % DO NOT CHANGE THIS AND DO NOT ADD ANY OPTIONS TO IT
\usepackage{algorithm}
\usepackage{algorithmic}

\usepackage{amsmath}
\usepackage{amssymb}
\DeclareMathOperator*{\argmaxA}{arg\,max} % Jan Hlavacek
\usepackage{subcaption}

\usepackage{newfloat}
\usepackage{listings}
\DeclareCaptionStyle{ruled}{labelfont=normalfont,labelsep=colon,strut=off} % DO NOT CHANGE THIS
\floatstyle{ruled}
\newfloat{listing}{tb}{lst}{}
\floatname{listing}{Listing}
\title{Predicting Bone Degradation Using Vision Transformer and Synthetic Cellular Microstructures Dataset}
\author{
    Mohammad Saber Hashemi\textsuperscript{\rm 1},
    Azadeh Sheidaei\textsuperscript{\rm *,1}
}
\affiliations{
    \textsuperscript{\rm *}Corresponding Author\\
    \textsuperscript{\rm 1}Iowa State University\\
    537 Bissell Rd, Ames, IA 50011 USA\\
    {mhashemi, sheidaei}@iastate.edu % jannesar
}

\begin{document}

\maketitle

\begin{abstract}
Bone degradation, especially for astronauts in microgravity conditions, is crucial for space exploration missions since the lower applied external forces accelerate the diminution in bone stiffness and strength substantially. Although existing computational models help us understand this phenomenon and possibly restrict its effect in the future, they are time-consuming to simulate the changes in the bones, not just the bone microstructures, of each individual in detail. In this study, a robust yet fast computational method to predict and visualize bone degradation has been developed. Our deep-learning method, TransVNet, can take in different 3D voxelized images and predict their evolution throughout months utilizing a hybrid 3D-CNN-VisionTransformer autoencoder architecture. Because of limited available experimental data and challenges of obtaining new samples, a digital twin dataset of diverse and initial bone-like microstructures was generated to train our TransVNet on the evolution of the 3D images through a previously developed degradation model for microgravity.
\end{abstract}

\begin{figure*}[t]
\centering
\includegraphics[width=1.0\textwidth]{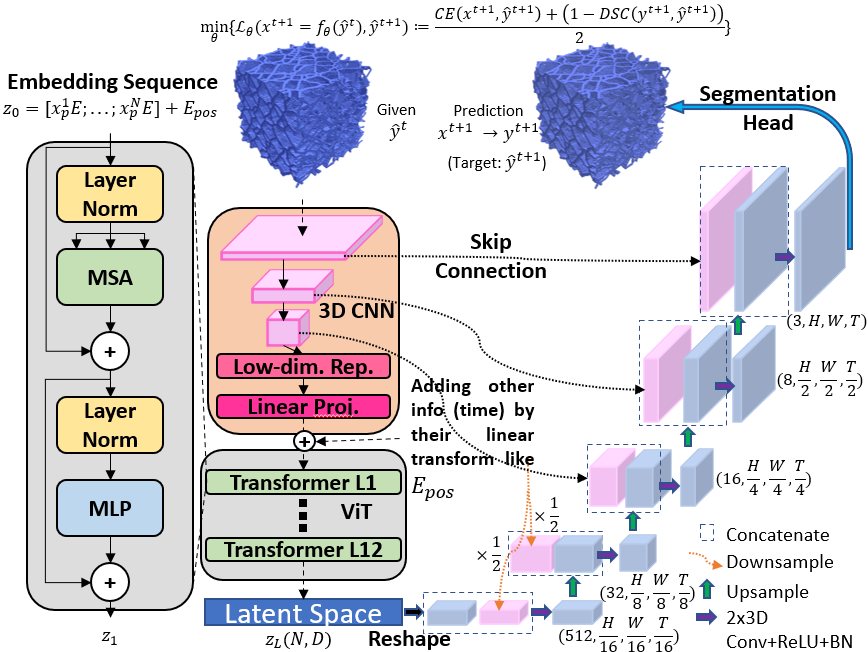} % Reduce the figure size so that it is slightly narrower than the column.
\caption{Overview of our proposed deep learning framework, TransVNet.}
\label{fig:TransVNet}
\end{figure*}

\section{Introduction}

Deep learning is a subfield of machine learning that uses artificial neural networks to model and solve complex problems. It is a type of machine learning that involves training artificial neural networks on large datasets to learn complex patterns and relationships in the data. Artificial neural networks have been devised to mimic human or natural intelligence with biological computing networks by encountering different input situations or samples and learning the correct actions, i.e., different neurons' activation, via optimizing an objective or loss function. Deep learners have outperformed not only traditional machine learning models but also humans in certain tasks, such as image recognition and game-playing, through customization and massive training for specific tasks \cite{janiesch2021machine}, thanks to the development of powerful hardware and efficient implementation of backpropagation algorithms \cite{rumelhart1986learning, wright2022deep}. However, there is still a long road to reach the ideal general intelligence with minimum data and computing necessities similar to humans. Deep learning has been used in many static or sequential applications, such as image recognition, speech recognition, natural language processing, and many other domains, such as drug discovery and genomics. Recently, Large Language Models (LLMs), such as GPT-3 (Generative Pre-trained Transformer 3) \cite{brown2020language}, BERT (Bidirectional Encoder Representations from Transformers) \cite{devlin2018bert}, and T5 (Text-to-Text Transfer Transformer) \cite{raffel2020exploring}, trained on massive amounts of text data available on the Internet, have demonstrated their outstanding capability in generating coherent and contextually relevant human-like texts that can be used for various applications like language translation, chatbots, and programming. Deep learners' capacity to learn complicated concepts at various levels of complexity is provided by layering several linear and non-linear elements for processing. They typically combine low-level features to obtain abstract high-level features, which can significantly alleviate the local minimum problem \cite{lecun2015deep, liu2021plant}.

Machine learning can be categorized in terms of the provided data features and learning methods as follows: Supervised, Semi-supervised, Self-supervised, and Unsupervised. Machine learning models can be either discriminative, such as logistic regression, support vector machines, decision trees, and random forests, or generative, such as Naive Bayes, Gaussian mixture models, Boltzmann machines, and deep generative models. Discriminative models learn to distinguish between different classes of data, thereby typically used for supervised learning tasks. In contrast, generative models learn to generate new data samples similar to the data they were trained on, thereby typically used for unsupervised learning tasks \cite{goodfellow2016deep}.

In this study, the specific application and the available data require a discriminative model to predict the next evolution of a given 3D segmented image, i.e., the microstructure, in the same given format, i.e., 3D segmented image. Since computational simulations were used for data generation, the dataset is automatically labeled, i.e., each given 3D input image is associated with a sequence of 3D images, and supervised learning is suitable. In the literature, different deep learning methods have been proposed for image sequence prediction, such as 3DCNN \cite{ji20123d} and ConvLSTM \cite{Azad_2019_ICCV}. However, our proposed method is tailored for 3D images and is simpler because each image only depends on the previous one; therefore, long-term memory issues do not affect our studied problem.

The quality of the bone texture depends on various factors, among which gravity is of great importance since it constantly applies a compressive force on body organs, especially bones. Under normal conditions on Earth, the compressive force helps in maintaining a healthy bone structure and preventing the weakening of bone. In the microgravity of outer space, bone cells react to the new loading condition quickly by degrading bone microstructures at a rate faster than normal. Despite the bone loss happening in the normal condition mostly due to aging and diseases, in lack of gravitational forces, the performance of bone cells significantly changes, and a high absorption rate causes a huge loss of bone minerals in a short time. Generally, the degradation rate is fast at first but then slows down. According to statistical studies, astronauts monthly lose about 2 percent of their bone mineral density in their lower limb bones and vertebrae as a consequence of being in microgravity and adapting to the new condition \cite{keyak2009reduction, vico2000effects, leblanc2000bone}. The bone loss graph in the absence of gravity looks like a curve with decreasing slope over time. According to \cite{qin2010challenges}, there are some similarities between microgravity bone degradation and bone loss due to aging in normal conditions. The research suggests that in both cases, the bone loss rate is rapid at the beginning of the degradation process, and the bone loss rate slows down until it reaches a steady state. The model presented by Qin predicts that a healthy person loses about 35 percent of their bone mineral density during a 3-year exposure to microgravity. Even though current spaceflights usually do not take longer than six months, prediction of changes in bone quality over a long time could come into use in understanding the mechanism of bone degradation. Moreover, future space flights may take longer, and more knowledge of the mechanism of bone resorption and formation will be beneficial to prevent bone loss in such situations.

Our virtually-generated dataset consists of the in-silico data of bone microstructure generation and its degradation throughout the study time (at most 36 months). The data format is 3D matrices or 3D voxelized images of the artificially generated and degraded bone microstructures under microgravity which is especially important for space exploration and experiment missions. Therefore, each data point is actually stored in a 4D matrix inside a ".mat" file as a result of the bone degradation code written in MATLAB software \cite{bagherian2020novel}. The fourth dimension is the month (e.g., index 0 denotes the initial 3D microstructure, index 1 denotes the changed/degraded microstructure after 1 month, ...). The 3D images are binary in the current dataset as there are two constituents inside the bones: soft bone marrow and hard bone mineral. Therefore, the features are segmented images, and the labels are the categorical values or classes of the voxels in 3D images. The input and output both have the same format, as we are interested in finding the degraded microstructure in detail, given the input 3D microstructure. However, the deep-learning framework is general for similar tasks. So far, we have generated more than 1000 microstructures with previously developed degradation model \cite{bagherian2020novel}. 10\% of the total dataset will be used for testing purposes, and the rest will be used for training. In the training cycles, 15\% of the available training data will used as the validation dataset.

Our deep learning algorithm is based on TransUNet \cite{chen2021transunet}, a framework which, itself, is based on U-Net \cite{ronneberger2015u} and Vision-Transformer \cite{dosovitskiy2020image}. However, our proposed model works with 3D images and predicts the next time frames for the degradation and similar time-series tasks. We utilize the Transformer's ability to embed mixed data sources. For instance, the time is embedded, as well as the embeddings related to the input microstructure or 3D image. The output is the microstructure at the next time step.

The outcomes of this computational framework include a novel virtual dataset of bone or cellular microstructures as well as their degradation evolution through a simulation method; and the novel deep learning framework, called TransVNet, that can be used for similar time-series prediction of 3D-image evolution and other generative models for material design; and some experimental results on how it performs given the available dataset. The generative variant is not part of this study. In Figure \ref{fig:TransVNet}, the high-level overview of our proposed algorithm is depicted.

\section{Methods and Materials}
\subsection{Virtual Data Generation}

\begin{figure}[h!tb]
    \centering
    \includegraphics[width=\linewidth]{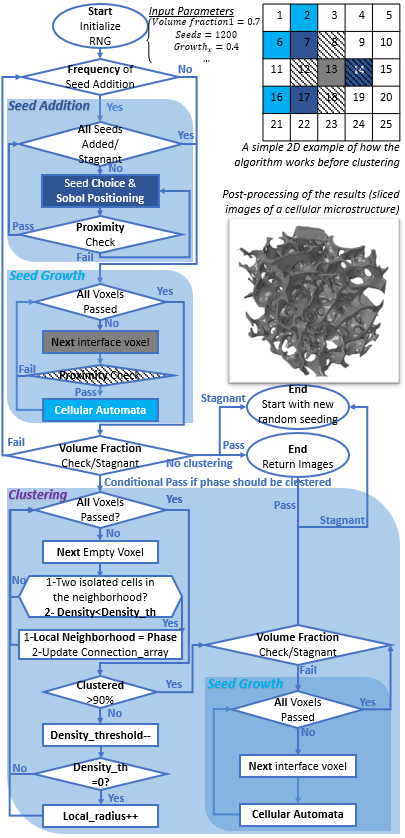}
    \caption{HetMiGen realization algorithm flowchart to generate initial 3D segmented microstructures.}
    \label{fig:HetMiGen}
\end{figure}

A fast and computationally efficient program called HetMiGen (short for Heterogeneous Microstructure Generator) has been developed to artificially generate 3D heterogeneous microstructures without the need to any reference 2D cut section images of physical material microstructures through expensive experimental methods such as SEM. This in-silico data generation is especially useful for computational materials science since it does not rely on the reconstruction techniques based on some expensive experiments, let alone the physical material processing complications with the shortcoming of not being able to consider all the design space for material design purposes. The C++ source codes can be compiled for and deployed on different machines with Linux or Windows OSs, and the executable can be readily run given a CSV file whose each line contains the microstructural parameters of the microstructure to be generated: the microstructure id number, the number of phases in addition to the background phase, the volume fraction of each phase, the number of initial seeds, the increment (positive or negative) of seeds’ addition in the next seed addition iterations, the frequency by which seeds are added, the radius of the local neighborhood to be checked for proximity for each phase (zero means no check is needed and seeds can be grown until they touch others), whether each phase should be clustered at the end, the rate of growth decay for each phase (zero means that growth parameters are fixed throughout the evolution iterations), and the probabilistic growth thresholds for Cellular Automate based on the considered neighborhood type (von Neumann in the current version of the codes). By changing the input parameters, it can generate multitudes of heterogenous microstructures of different material systems consisting of two or more material phases with different morphologies. The parameters have an affinity to the physical process of manufacturing and the thermomechanical evolution of microstructure as well. The details of this algorithm is discussed in a paper to be published by the author. A sufficiently large number of cellular microstructures were generated using this code to create the initial dataset of 3D segmented bone-like images. The overview of the algorithm flowchart is shown in Figure \ref{fig:HetMiGen}.

Based on the initial dataset, the degraded or time-series evolution of the microstructures has been simulated using a previously developed method and its associated parallel MATLAB code \cite{bagherian2020novel}. Those data samples are then preprocessed in MATLAB such that the high-quality ones with more than 95\% clustering in the bone mineral segment are filtered, and an equal number of microstructures for different volume fraction ranges are present since it is the most important feature of the 3D images, and a balanced dataset is needed for accurate prediction after training. Therefore, data sampling is balanced among different volume fraction ranges during the training. In addition to this balanced sampling, appropriate data augmentation techniques, such as random rotation and random flipping, are considered during training for robustness. The main features considered are the segmented 3D images with their associated time stamps. However, other expert-picked features of the input microstructures, such as their TPCF and TPCCF statistical descriptors, can also be considered to supplement the aforementioned main features if they are needed or proven to improve the final network performance.

\subsection{TarnsVNet, a Hybrid Autoencoder}

To the author's knowledge, the 3D image sequence prediction problem has not been studied extensively yet. However, our proposed method, TransVNet, is based on the state-of-the-art in computer vision, as it can take advantage of local convolutional operators and global attention ones. More specifically, our network uses local and global features in encoding the input data and uses the localized information of the previous or input microstructures in reconstructing or decoding the encoded features via skip connections to associated CNN-extracted features of the input image. Another advantage of our proposed method compared to 2D networks such as TranUNet is that it directly processes 3D images. One might argue that it is possible to consider a 2D network operating sequentially on a stack of 2D images that forms the input 3D image together. Nevertheless, such an approach discards the 3D information and relationships in the input image and may not be as accurate. This is especially important when the input image has long 3D features in all directions, which is the case for cellular morphologies found in nature, such as bone microstructures. Due to the much bigger input feature size of 3D images than 2D images and limited computational resources, a minimum number of features for different levels of encoder and decoder has been considered. Also, the ViT parameters were fixed for faster training and lower computational cost. However, no pre-trained 3D CNN feature extractor was found, so the CNN feature extractor is completely trained. As the objective is the correct labeling of the output voxels for segmentation or classification, the Cross-Entropy (CE) loss criterion was considered in the loss function by applying the Softmax function on the segmentation mask, i.e., the log-likelihoods of possible classes of data which are binary in our studied case. As many segmentation studies have reported DSC scores for their performance measurement, it is also used to assess the accuracy of the trained network and complement the Cross-Entropy term in the loss function. The DSC measures the spatial overlap between two segmented images, A and B target regions, and is defined as $DSC(A,B)= 2(A\bigcap B)/(A+B)$. In binary manual segmentation, this coefficient may be derived from a two-by-two contingency table of segmentation classification probabilities. Thus, minimizing the total loss functional $\mathcal{L}$ with respect to the network parameters, $\theta$, is the learning objective:

\begin{multline}
% \begin{split}
    \mathcal{L}^{b}_{\theta} \left[ x^{t+1}=f_{\theta}(\hat{y}^{t}),\hat{y}^{t+1} \right] \\= \frac{CE(x^{t+1},\hat{y}^{t+1})+(1-DSC(y^{t+1},\hat{y}^{t+1}))}{2} \\= \frac{1}{2}\Biggr[ \left( \frac{-1}{H\times W \times T}\sum_{n=1}^{H\times W \times T} log(\frac{exp(x_{n,\hat{y}_{n}^{t+1}}^{t+1})}{\sum_{c=1}^{C} exp(x_{n,c}^{t+1})}) \right)\\+ \left( 1-DSC(\argmaxA_c\{\frac{exp(x_{n,c}^{t+1})}{\sum_{c=1}^{C} exp(x_{n,c}^{t+1})}\},\hat{y}^{t+1}) \right) \Biggr]
% \end{split}
\label{eq:1}
\end{multline}
, where $b$ denotes sample $b$ in a batch of size $B$, $\hat{y}^{t}$ is the input of the network (the segmented image at time step t), $\hat{y}^{t+1}$ is the target of the network (the segmented image at time step t+1), $x^{t+1}$ is the output of the segmentation mask layer (the log-likelihoods of each class at different voxels), $y^{t+1}$ is the label or segmented image associated with $x^{t+1}$, $C$ is the number of classes (2 in our case), and $N$ is the total number of voxels.  

As seen in the overview of the architecture, the encoder is hybrid, meaning it consists of a 3D CNN feature extractor followed by a vision transformer. The goal is to predict the next evolution of the microstructure, called $\hat{y}$, with $\hat{C}$ channels for the desired number of segmentation classes (two in this study because each voxel represents the material phase at that location, which can be either bone marrow or bone mineral), given a 3D input image $y \in \mathbb{R}^{H \times W \times T}$ ($160^{3}$ dimensions in this study following resizing of the HetMiGen outputs of $150^{3}$ dimensions) with three default input channels (the RGB format in general). The CNN local feature extractor encodes the input into its low-dimensional high-level features, with $C^{'}$ (32 in this study) channels and a dimension size of $H^{'}=\frac{H}{2^{\#CNNDownScaling}}=\frac{160}{2^{3}}$ (in this study). Each block consists of a double 3D Convolution followed by ReLU nonlinearity and Batch Normalization layers, and a DownSampler through $1 \times 1 \times 1$ Convolution with a stride of two. The vision transformer needs a series given by image sequentialization. Therefore, the input image is tokenized into 3D flattened patches of size $P$:

\begin{equation}
    \{x_{i}^{P} \in \mathbb{R}^{P^{3}.C}|i=1,...,N=\frac{HWT}{P^{3}}\}
\end{equation}
Then, these patches will be transformed into D-dimension vectors, with D being the hidden state dimension of the vision transformer, by a learnable linear transformation, $E\in R^{(P^{3}.C^{'})\times D}$. The important point is that we also add the patch positional embedding and the time embedding, via a linear transformation, to the image embedding to encode features not present in the raw input image. Then, the transformer with $L$ layers will transform the embedded sequence as follows (MSA is Multi-head Self Attention, MLP is multi-layer perception, and LN is layer normalization operator).
\begin{equation}
    z_0 =[x_1^{P}E;...;x_N^{P}E]+Emb_{time}+Emb_{position}
\end{equation}
\begin{equation}
    z_l^{'} = MSA(LN(z_{l-1}))+z_{l-1}
\end{equation}
\begin{equation}
    z_l = MLP(LN(z_l^{'}))+z_l^{'}
\end{equation}

The decoder is a 3D CNN Upsampler. There are multiple cascaded trainable blocks to reach the high-dimensional output or segmentation mask from the low-dimensional encoded sequence of feature vectors. First, the sequence of $N$ encoded features, $z_{L} \in \mathbb{R}^{N \times D}$, is reshaped into a 3D image with the dimension of $D \times \frac{H^{'}}{P} \times \frac{W^{'}}{P} \times \frac{T^{'}}{P}$. Then, each upsampling block will double the size of the image. Each block consists of a trilinear upsampling function with an enlargement factor of two, a double 3D-Convolution, a ReLU nonlinearity, and a Batch Normalization. As mentioned above, the V-shaped architecture enables feature aggregation via Skip Connections from the CNN encoder layers to their associated CNN decoder ones. The decoder layers with input sizes lower than the smallest CNN encoder features can get their encoder features via Max Pooling of the last CNN encoder outputs, i.e., the smallest features.

\section{Results}
\subsection{Data generation}
\begin{figure}[h]
    \centering
    \includegraphics[width=\linewidth]{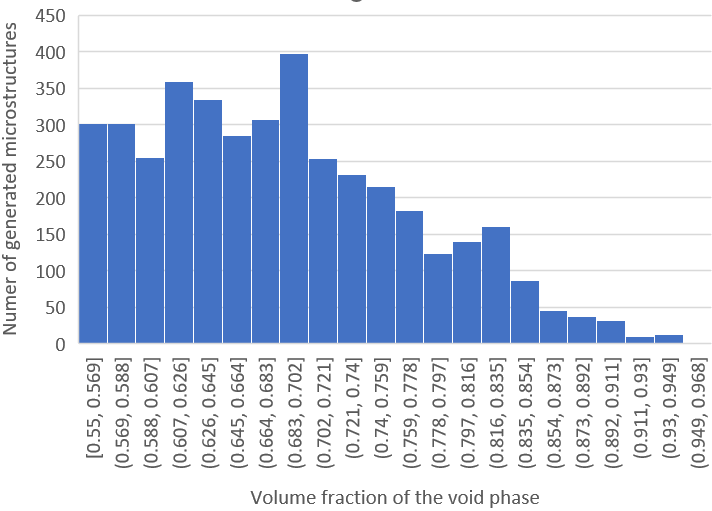}
    \caption{Histogram of the number of microstructures generated versus the volume fraction ranges.}
    \label{fig:Histogram}
\end{figure}

Using the HetMiGen code, a diverse dataset of initial bone-like geometries represented by binary-segmented 3D images was generated. Then, based on the previous study, we simulated the degradation of the microstructures during 36 months of time steps. This provided us with a sizeable training dataset to test the performance of TransVNet. The result of data generation is summarized by the histogram of Figure \ref{fig:Histogram}. As inferred from the figure, the dataset is unbalanced with respect to the volume fraction of the bone marrow segment in the initial microstructures. Therefore, the high-volume fraction microstructures were sampled more frequently in training in addition to data augmentation through rotation and flipping.

\subsection{The network performance and ablation study}

\begin{figure}[h]
    \centering
    \includegraphics[width=\linewidth]{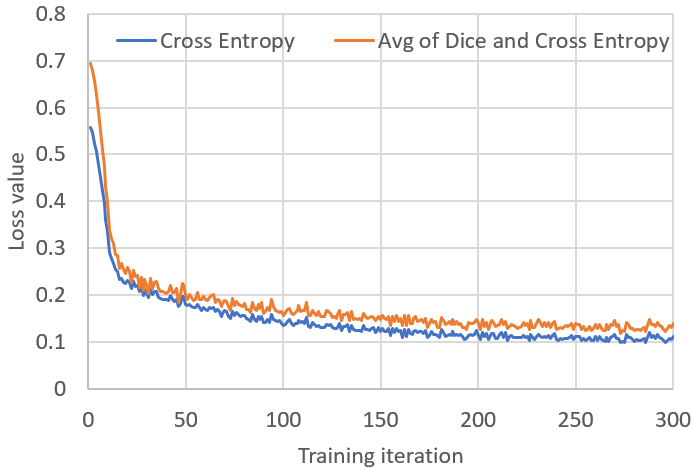}
    \caption{The loss values during the training; red line denotes the main training signal, the average of the Dice loss and the Cross-Entropy loss, while the blue one shows the Cross-Entropy alone.}
    \label{fig:Training Loss}
\end{figure}

To the best of our knowledge, there is no appropriate baseline method for our specific problem, the time-series prediction of 3D segmented images. However, the initial training performance is better than that of TransUNet applied to 2D medical segmentation tasks. The curves of the loss values during the training are plotted in figure \ref{fig:Training Loss}. The main and only training signal is the average value of the Dice loss and the Cross-Entropy loss shown by the red line. The Dice loss is calculated as $Dice = 1 - DSC$ as explained above. Its value at the end is 0.084, which means that DSC is 91.64\%. The best DSC score of TransUNet was 89.71\%, calculated as the average value for different classes of organs segmented by their network. We also considered Hausdorff distance as another performance metric in our experiments. Hausdorff distance (HD) measures how far two subsets of a metric space are from each other.

\begin{figure*}[h]
\centering
\begin{subfigure}[t]{0.8\textwidth}
    \includegraphics[width=\linewidth]{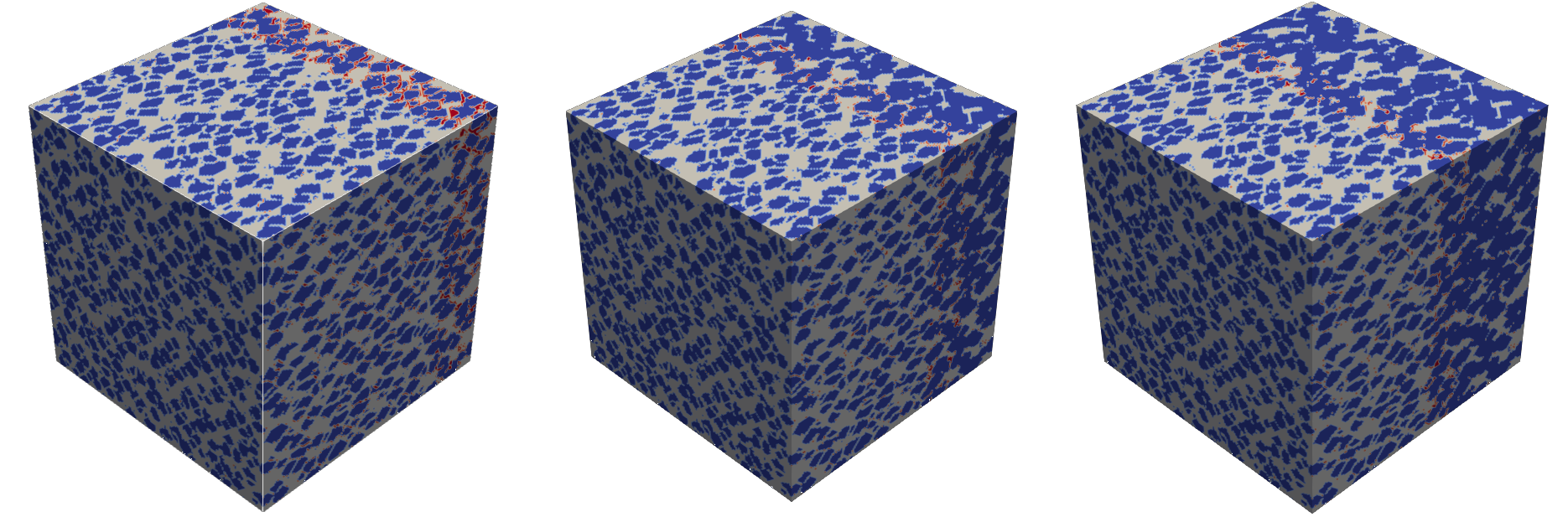}
    \caption{Evolution of microstructure 025980 with an initial mineral volume fraction of 0.406.}
\end{subfigure} %\hspace{\fill} % maximize horizontal separation
\begin{subfigure}[t]{0.8\textwidth}
    \includegraphics[width=\linewidth]{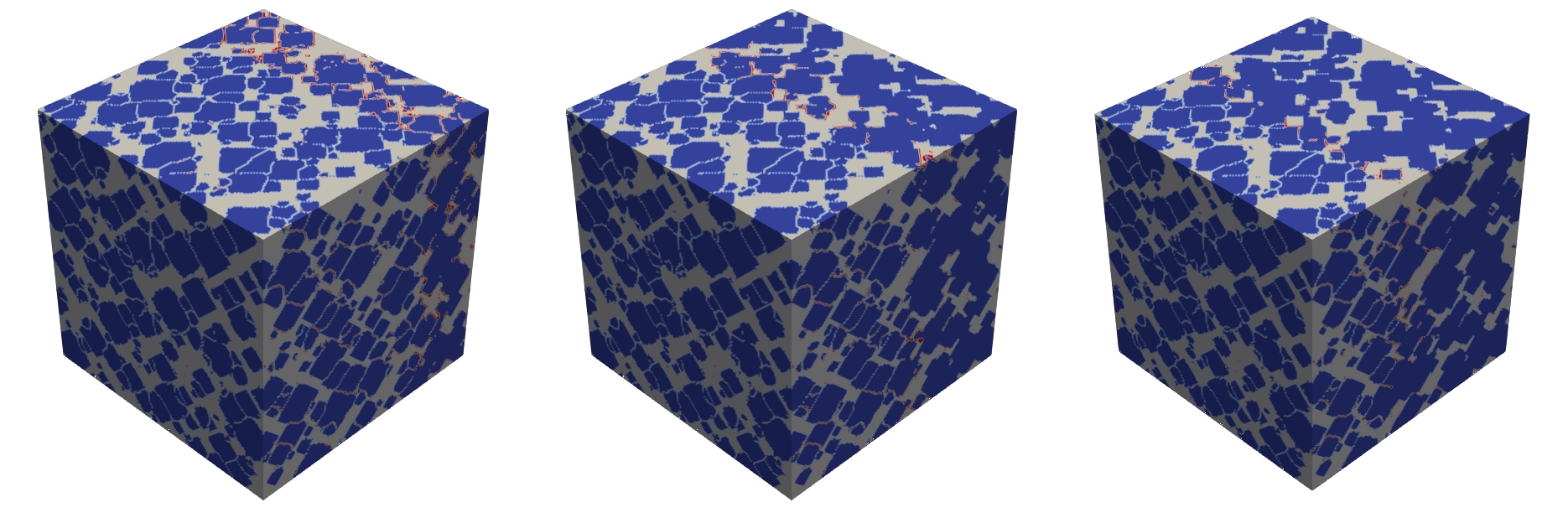}
    \caption{Evolution of microstructure 101962 with an initial mineral volume fraction of 0.235.}
\end{subfigure}
\caption{Samples of the network performance on the test set (leftmost column: evolution after 4 months; middle column: evolution after 8 months; rightmost column: evolution after 12 months) (red color: network error).}
\label{fig:vis}
\end{figure*}

\begin{table*}[h!tb] \centering
\caption{Ablation and scaling experimental results on our test dataset.}
\label{experiment}
% Use: \begin{tabular{|lcc|} to put table in a box
\begin{tabular}{||lcc||} \hline
\textbf{Method} & \textbf{DSC} ($\in [0, 1] \uparrow$) & \textbf{HD} (mm or voxels $\downarrow$) \\ \hline
ViTOnly-pretrained & 0.040122 & 14.725012 \\
ViTOnly & 0.118725 & 21.709163 \\
ViTOnly-patch8 & 0.208199 & 20.731199 \\
Hybrid-patch8 & 0.981572 & 0.033904 \\
Hybrid-patch8-months4 & 0.960275 & 0.300317 \\
Hybrid-resolution160-patch2-months4 & 0.978929 & 0.184443 \\
Hybrid-resolution160-patch2-months4-epoch1 & 0.955390 & 0.722508 \\ \hline
\end{tabular}
\end{table*}

We performed an ablation study to further assess the capabilities of our network and show the superiority of our proposed TransVNet architecture compared with other possible models. To make the comparison faster and more computationally efficient, we first considered resized microstructures with $64^3$ resolution instead of the original $150^3$ resolution in our generated dataset. The first model on the encoder side was simply the ViT trained on ImageNet with fixed parameters in its layers and trainable ones in the embedding layer with a patch size of 16, thereby having $\frac{64}{16} \times \frac{64}{16} \times \frac{64}{16} = 64$ tokens for global information inference through the transformer-based architecture. After four training epochs, the results showed very poor performance on the unseen test dataset. Making the hidden layers trainable also did not improve the performance to a satisfactory level; however, considering the lower patch size of 8, leading to the higher number of tokens as the transformer layers' input, improved it significantly yet less than a satisfactory level. Then, we considered our proposed hybrid model in the encoder with Skip Connections to the decoder, which resulted in very high and satisfactory performance metrics. To determine whether our proposed method is also efficient in the more challenging task of next-four-step prediction of the segmented image with morphological changes much more than those in the next-step prediction, the target segmented image after four months was considered in training, i.e., replacing $\hat{y}^{t+1}$ with $\hat{y}^{t+4}$ in Equation \ref{eq:1}. The performance metrics show that it is very accurate in the new data representation yet performs at levels lower than the less challenging task. 

Next, we focused on the scalability of our proposed method by considering a network with an architecture similar to the last successful model yet suitable for the original-resolution inputs and outputs; and by starting the training process from the shared learned parameters of the trained lower-resolution network. This experiment also resulted in high performance metrics. Therefore, our proposed method is advantageous since it can achieve satisfactory performance, even after one training epoch, when the learned weights of the lower-resolution network are transferred for fine-tuning a high-resolution network. The detailed numerical results are provided in Table \ref{experiment}. Sample performance results of the last network on the test dataset are visualized in Figure \ref{fig:vis}. The white color shows the mineral phase of the 3D bone microstructures while the blue one denotes the marrow phase. The red voxels show the discrepancy between the network prediction and the ground truth, i.e., the bone degradation simulation results. The leftmost, middle, and rightmost columns demonstrate the fourth-month, eighth-month, and twelfth-month snapshots of the evolution, respectively.

\section{Discussion and Conclusions}

In conclusion, our scalable and fast HetMiGen code was used to generate an artificial dataset of 1000 different geometries of the porous bone microstructures at their initial time step. Then, the time-series evolution of the initial 3D segmented images was simulated via a previously developed bone degradation code. We proposed a new deep learning network, called TransVNet, which was directly applied to the fast prediction of the bone degradation task given our artificially created (in-silico) dataset. Our experimental results show that TransVNet has a superior performance compared with its ancestor, TransUNet, which has been applied to the different yet less challenging task of 2D medical image segmentation, and our proposed method is scalable as it remains very accurate even for image sequences with higher levels of changes and as it can be efficiently fine-tuned for high-resolution prediction using previously trained parameters of a lower-resolution network. As for future studies, the effect of different hyperparameters of the current model on its performance can be further studied. Furthermore, the architecture can be changed by considering Transposed Convolution operators in the decoder and processing more than one input at a time step using the transformer capability in finding sequential relationships, especially for other time-series data with longer history dependency. The results of such variations can be compared to determine the best-performing network.

% \section{Acknowledgments}

% The authors acknowledge the funding support provided by ISSNL and greatly appreciate the assistance of Mr. Amirhossein Bagherian for using their previously-developed bone degradation code to complete the data generation for this study.

% \bigskip
% \noindent Thank you for reading these instructions carefully. We look forward to receiving your electronic files!

\bibliography{main}

\end{document}